\documentclass[a4paper,nofootinbib,preprintnumbers,floatfix,superscriptaddress,pra,twocolumn]{revtex4-1}
\usepackage{amsmath, amsthm, amssymb, amsfonts}
\usepackage{graphicx}
\usepackage{dcolumn}
\usepackage{bm,bbm}
\usepackage{color}
\usepackage{hyperref}

\newcommand{\eqnref}[1]{(\ref{#1})}
\newcommand{\figref}[1]{Fig.~\ref{#1}}

\newcommand{\ah}{\hat a}

\newcommand{\half}{\frac{1}{2}}                         
\newcommand{\vacr}{|vac\rangle}
\newcommand{\vacl}{\langle vac|}
\newcommand{\bra}[1]{\langle #1|}
\newcommand{\ket}[1]{|#1\rangle}

\newcommand{\ketbra}[3][]{\ket{#2}_{#1}\!\bra{#3}}      
\newcommand{\ceta}{\tilde{\eta}}						

\newcommand{\ba}{\begin{eqnarray}}
\newcommand{\be}{\begin{equation}}
\newcommand{\ee}{\end{equation}}

\newcommand{\ea}{\end{eqnarray}}
\newcommand{\ban}{\begin{eqnarray*}}
\newcommand{\ean}{\end{eqnarray*}}

\DeclareGraphicsExtensions{.pdf,.png,.jpg}

\makeindex

\begin{document}

\title{Bell tests for continuous variable systems using hybrid measurements\\ and heralded amplifiers}
\date{\today}

\author{Jonatan Bohr Brask}\affiliation{ICFO-Institut de Ciencies Fotoniques, Av. Carl Friedrich Gauss 3, 08860 Castelldefels (Barcelona) - Spain}
\author{Nicolas Brunner}\affiliation{H.H. Wills Physics Laboratory, University of Bristol, Tyndall Avenue, Bristol, BS8 1TL, UK}
\author{Daniel Cavalcanti}\affiliation{Centre for Quantum Technologies, National University of Singapore, 3 Science drive 2, 117543 Singapore}
\author{Anthony Leverrier}\affiliation{ICFO-Institut de Ciencies Fotoniques, Av. Carl Friedrich Gauss 3, 08860 Castelldefels (Barcelona) - Spain}\affiliation{Institute for Theoretical Physics, ETH Zurich, 8093 Zurich, Switzerland}

\begin{abstract}
We present Bell tests for optical continuous variable systems, combining both hybrid measurements (i.e. measuring both particle and wave aspects of light) and heralded amplifiers. We discuss two types of schemes, in which the amplifier is located either at the source, or at the parties' laboratories. The inclusion of amplifiers helps to reduce the detrimental effect of losses in the setup. In particular, we show that the requirements in terms of detection efficiency and transmission losses are significantly reduced, approaching the experimentally accessible regime.
\end{abstract}


\maketitle

\section{Introduction}
\label{sec.intro}

Quantum mechanics predicts that separated observers sharing entanglement may observe correlations which cannot be explained by any classical mechanism. Signalling between the parties is excluded by ensuring space-like separation such that the signal would have to travel faster than light. Moreover, an explanation based on a classical strategy using a common source can also be ruled out through the violation of Bell inequalities \cite{bell1964}. This remarkable phenomenon known as quantum nonlocality, which has been the root of much debate, is today actively investigated in order to get a deeper understanding of the foundations of quantum mechanics. Its study is further motivated by the development of novel quantum technologies. Nonlocality is recognized as a powerful resource for information processing. It allows for the reduction of communication complexity \cite{buhrman2010}, and plays a central role in device-independent protocols, which range from cryptographic primitives such as quantum key distribution \cite{acin2007,hanggi2010,masanes2011} and randomness expansion \cite{pironio2010,colbeck2011}, to quantum state estimation \cite{bardyn2009,bancal2011,rabelo2011}.

Experimental evidence for quantum nonlocality is compelling \cite{aspect1999}. All tests of nonlocality -- known as Bell tests -- performed so far have confirmed quantum predictions. Nevertheless, none of these experiments could yet conclusively rule out local models, since they all suffer from at least one out of two loopholes. First, the \emph{locality loophole} is opened whenever the measurement events of remote parties are not space-like separated; a classical model using subluminal communication can then in principle explain the data. This loophole can be closed in photonic experiments, since photons can be transmitted over large distances with relative ease and measured with fast detectors. However, optical Bell tests have been hampered by the so-called \emph{detection loophole} \cite{pearle1970}. When the detection efficiency is below a certain threshold, it becomes possible to reproduce experimental data with a local model. Typical photo-detection efficiencies (including transmission losses and detector efficiency) in optical Bell tests are currently too low for closing the detection loophole. Conversely, Bell tests carried out with entangled atoms have closed the detection loophole \cite{rowe2001}, while leaving the locality loophole open due to the relatively slow atomic measurements.

Thus, while both loopholes have been closed experimentally, a definitive loophole-free Bell test, in which both loopholes would be closed simultaneously, is still missing. Such an experiment, much sought after, would not only be central from the perspective of the foundation of mechanics, but would also represent a crucial step towards the implementation of device-independent applications, which is still highly challenging \cite{gisin2010}.

In this context, it is interesting to consider optical continuous variable systems \cite{gilchrist1998}. Here the main advantage comes from the high efficiency at which homodyne measurements can be implemented. Bell tests based on continuous variables have until recently proven to be illusive, either requiring states (or measurements) which cannot be achieved with current technology \cite{banaszek1998}, or yielding very small violations \cite{garcia-patron2004,nha2004}. 

Recently however, Cavalcanti et al. \cite{cavalcanti2011} showed that a scheme based on hybrid measurements -- that is, combining both homodyne and photodetection measurements -- can lead to large Bell inequality violations for continuous variable systems. Moreover, the required states and measurements can be implemented experimentally. In this scheme, the thresholds for a loophole-free Bell test in terms of transmission and detection efficiencies are comparable with those of the best schemes known for discrete systems \cite{vertesi2010}. Additionally, it has been shown that continuous-variable Bell tests based on hybrid measurements can lead to arbitrarily low detection-efficiency thresholds when considering more complex (experimentally unfeasible) states \cite{quintino2011,araujo2011}. Finally, the idea of hybrid Bell tests was also demonstrated to be relevant in the context of atom-photon systems \cite{sangouard2011,chaves2011}, where the photodetector efficiency threshold is reduced due to the high atomic detection efficiency \cite{brunner2007,cabello2007}.

Here we propose to combine hybrid Bell tests with heralded amplifiers \cite{ralph2009,marek2010}. The latter were recently discovered to allow for noiseless amplification of small coherent states of light, and have been experimentally demonstrated \cite{xiang2010,ferreyrol2010,zavatta2010,usuga2010}. Such a noiseless amplification process is necessarily probabilistic -- a deterministic implementation is impossible, due to the linearity of quantum mechanics -- but can be achieved in a heralded way. We show that when heralded amplifiers are included in hybrid Bell tests, both the transmission and detection efficiencies required for a loophole-free experiment can be significantly reduced compared to the scheme of Ref. \cite{cavalcanti2011}, approaching the currently experimentally accessible regime. The main idea behind our schemes is that the amplification counteracts the detrimental effect of photon losses. First we consider a scheme in which the amplifier is placed at the source (see Fig.~1a), enabling generation of states that are more robust to losses. Second, we discuss a setup in which the amplifiers are located just before the local measurements of the parties (see Fig.~1b), hence acting as local filters, heralding the presence of a signal.

\section{Setup and optimal states}
\label{sec.meas_optstat}

\begin{figure}
\includegraphics[width=.49\textwidth]{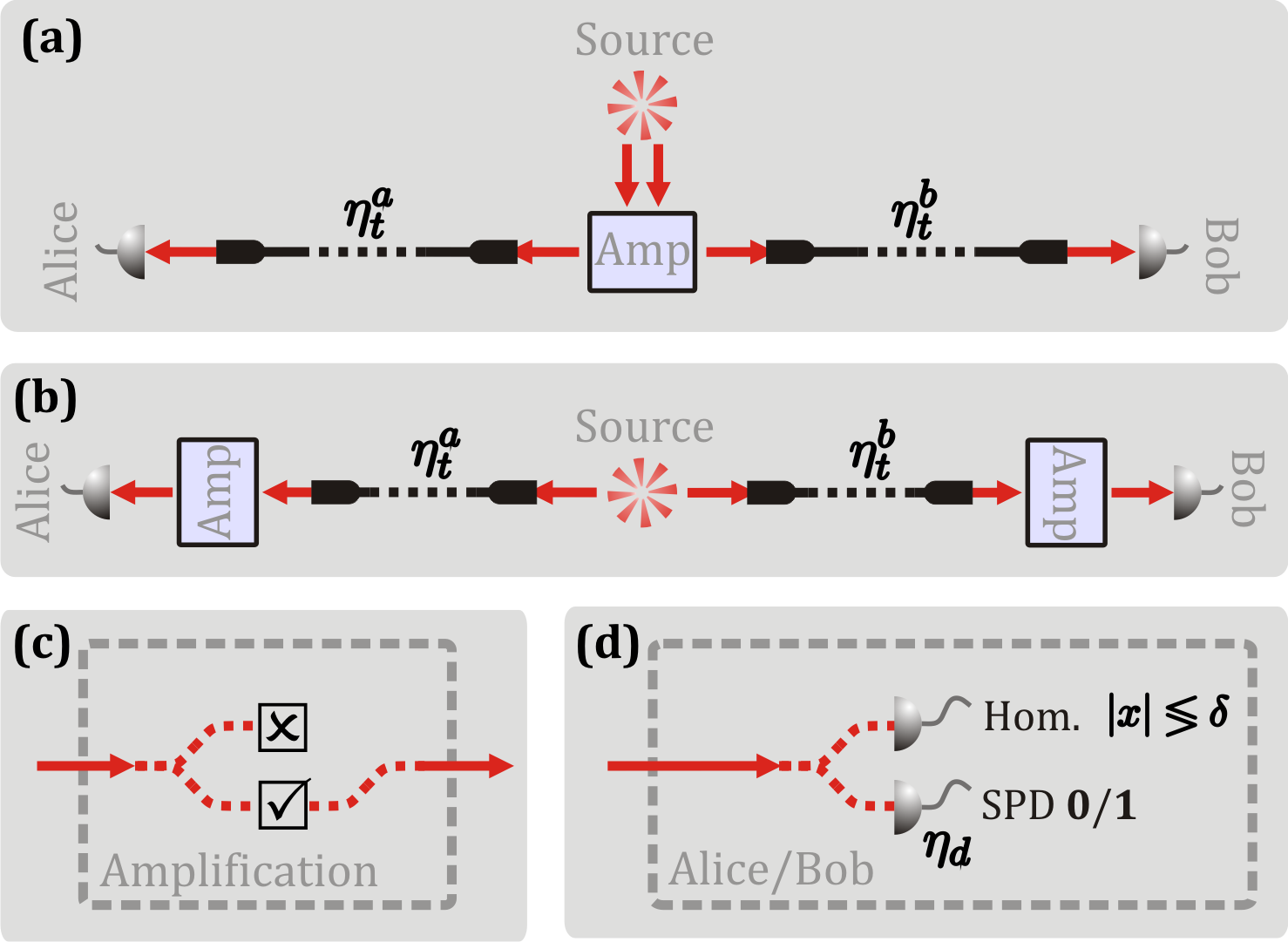}
\caption{(Colour online) \textbf{(a)} Bell test with amplification at the source. The amplifier enhances the entanglement produced by the source, thus making it more robust against losses ($\eta_t^a$ and $\eta_t^b$ are the transmission of the channels). \textbf{(b)} Bell test with local amplification by the parties. Before choosing a measurement basis, Alice and Bob locally filter their incoming signal using an amplifier \textbf{(c)} The amplification is a probabilistic, but heralded process. The Bell protocol only proceeds when amplification is successful. \textbf{(d)} Hybrid measurements. Each party can choose between photo-detection or homodyning. Single-photon detectors (SPD) have efficiency $\eta_d$. The continuous outcome of the homodyne measurement is binarised using a binning.}\label{fig.setup}
\end{figure}

The general setup we consider is outlined in \figref{fig.setup}. A source prepares an entangled state of two light modes. Each mode is transmitted to one of the parties, Alice and Bob. Heralded amplification may be applied either at the source, or by each party locally. Throughout the paper, we will mainly focus on the Clauser-Horne-Shimony-Holt (CHSH) Bell inequality \cite{chsh}. Each party then has the choice between two measurements, yielding binary outcomes. We consider hybrid measurements (see Fig.~1d), meaning that Alice and Bob can perform either a photodetection measurement ($\hat{N}$), or a homodyne quadrature measurement ($\hat{X}$). For the photo-detection measurement, we consider detectors that simply determine the presence (or absence) of light. No detectors resolving the number of incident photons are required. The result of the measurement is labelled by $+1$ when the detector clicks, and $-1$ otherwise. The quadrature measurement is fixed to be along the phase-space $X$-axis. The output of this measurement, which is continuous yielding a real number $x$, is then binarized: the outcome is labelled $+1$ if $|x| > \delta$ and $-1$ if $|x| \leq \delta$, where $\delta$ is a constant to be optimized.

We will investigate the performance of various entangled states of light for the hybrid Bell scenario described above. Our central concern will be robustness to transmission losses and limited detection efficiency. Hence, it will be useful to define an effective Bell operator, taking losses into account. Both measurements, $\hat{X}$ and $\hat{N}$, can be conveniently represented by Positive-Operator-Valued-Measures (POVMs). A photodectector with limited effieciency $\eta_d$ is modeled by an ideal (perfectly efficient) detector preceded by a fictitious beam-splitter with transmittivity $\eta_d$. The POVM element corresponding to no click is given by (in the Fock-state basis)
\begin{equation}
\hat{N}_0 = \vacr\vacl + \sum_{i=1}^\infty (1-\eta_d)^i \ket{i}\bra{i} , \\
\end{equation}
and the complementary element is $\hat{N}_1 = \mathbbm{1} - \hat{N}_0$. The measurement operator corresponding to photo-detection is then given by $\hat{N} = \hat{N}_1 - \hat{N}_0$.  For a perfectly efficient homodyne measurement followed by our binarisation process, the projector corresponding to the outcome $+1$ is given by
\begin{equation}
\hat{X}_{<}^{id} = \int_{-\delta}^{\delta} dx \ket{x}\bra{x} ,
\end{equation}
where $\ket{x}$ denotes an eigenstate of the quadrature operator. The corresponding POVM element for a lossy homodyne measurement with efficiency $\eta_h$ can be expressed in the Fock-state basis as
\begin{equation}
(\hat{X}_{<})_{nm}  = \sum_{k=0}^{\min n,m} (1-\eta_h)^k \eta_h^{\half(n+m)-k} \nu^k_{nm} (X_{<}^{id})_{nm}
\end{equation}
where $\nu^k_{nm}  = \sqrt{ \binom{n}{k} \binom{m}{k} }$ and $\binom{n}{k}$ is a binomial coefficient. The complementary POVM element is given by $\hat{X}_{>} = \mathbbm{1} - \hat{X}_{<}$, and the measurement operator is $\hat{X} = \hat{X}_{>} - \hat{X}_{<}$. Transmission losses, in the channel linking the source to the local measurements, can be accounted for by multiplying both $\eta_d$ and $\eta_h$ by an additional factor $\eta_t$ (effectively adjusting the transmission of the fictitious beam splitters). With the above measurement operators, the CHSH inequality can now be expressed as $|\langle \hat{B} \rangle| \leq 2$, where the Bell operator is given by
\begin{equation}
\hat{B} = \hat{X} \otimes (\hat{X} + \hat{N}) + \hat{N} \otimes (\hat{X} - \hat{N}) .
\end{equation}
Since $\hat{B}$ depends on the loss parameters $\eta_d$, $\eta_h$, and $\eta_t$, by diagonalizing $\hat{B}$ we can find the entangled state that achieves the largest CHSH violation for a given amount of losses. However, this approach can only be implemented efficiently if we consider specific subspaces with restricted photon numbers \cite{quintino2011}. Here we restrict our attention to the subspace featuring at most 2 photons in total. We assume perfectly efficient homodyne measurements, $\eta_h = 1$, which is a good approximation in certain experimental setups. 

We obtain the optimal states and minimal efficiencies for two scenarios. First, a symmetric scenario, in which the source is located halfway between Alice and Bob, hence leading to $\eta_t^a=\eta_t^b=\eta_t$. The threshold efficiencies are found to be 
\begin{align}
\ceta_t = 80.5\% \hspace{0.6cm} (\eta_d=1) , \\
\ceta_d = 64.8\% \hspace{0.6cm} (\eta_t=1) ,
\end{align}
and the corresponding optimal state are
\begin{align}
\label{eq.optstate_symt}
\ket{\psi^{\mathrm{sym}}_{t}} = 0.18\ket{00} - 0.70(\ket{20} + \ket{02}) , \\
\label{eq.optstate_symd}
\ket{\psi^{\mathrm{sym}}_{d}} = 0.22\ket{00} - 0.69(\ket{20} + \ket{02}) .
\end{align}
Both threshold values are significantly lower than in the original scheme of Ref.~\cite{cavalcanti2011} (given by $\eta_d=71.1\%$ and $\eta_t=84\%$), where the entangled state considered was of the form $\ket{20} + \ket{02}$. Second, we consider an asymmetric scenario, in which the source is located next to Alice's lab, hence $\eta_t^a=1$, $\eta_t^b=\eta_t$. The thresholds efficiencies are
\begin{align}
\label{eq.optthr_asymt}
\ceta_t = 66.7\% \hspace{0.6cm} (\eta_d=1) , \\
\ceta_d = 64.8\% \hspace{0.6cm} (\eta_t=1) ,
\end{align}
The optimal state with respect to photo-detection efficiency is again $\ket{\psi^{\mathrm{sym}}_{d}}$, whereas for transmission we have
\begin{equation}
\label{eq.optstate_asymt}
\ket{\psi^{\mathrm{asym}}_{t}} = 0.13\ket{00} - 0.86\ket{20} - 0.49\ket{02}.
\end{equation}
Thus, both the symmetric and asymmetric schemes offer a significant improvement in terms of loss tolerance over previous proposals. This naturally raises the question of whether the required states can be implemented experimentally, which is the subject of the next section.

\begin{figure}
\includegraphics[width=.48\textwidth]{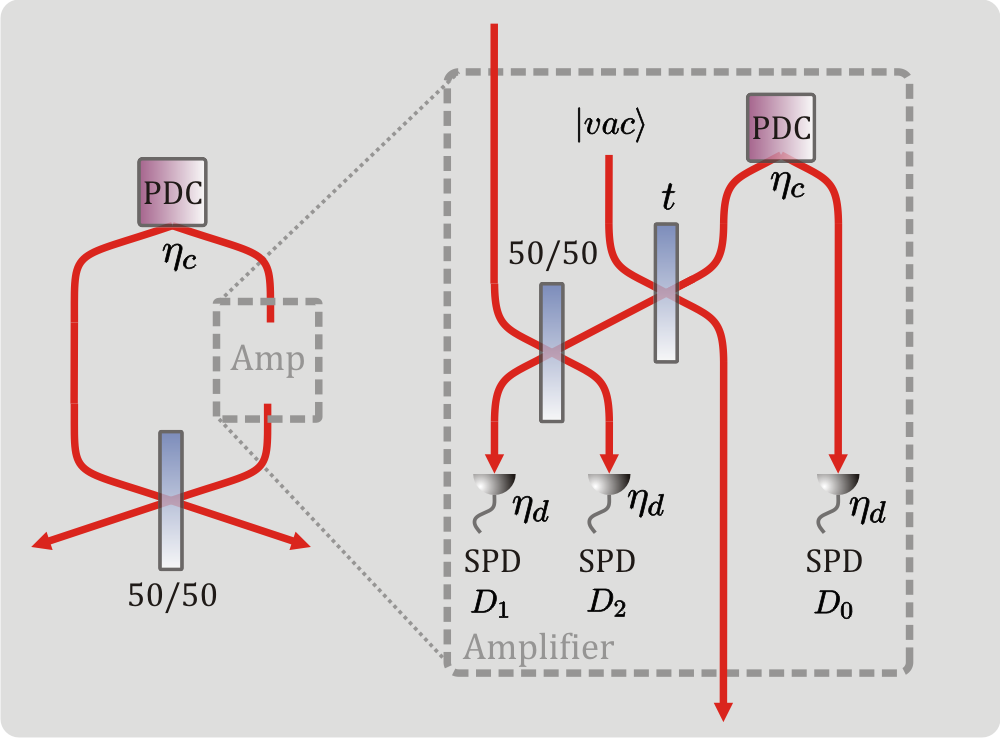}
\caption{(Colour online) Source with amplification. The source consists of a PDC source producing a two-mode squeezed vacuum state, and a heralded amplifier. The amplifier uses a single photon ancilla, created by PDC followed by detector $D_0$, which is then mixed with the signal to be amplified. Successful amplification requires a click in $D_0$ and either $D_1$ or $D_2$. By adjusting the strength of the squeezing $\lambda$ and the gain of the amplifier $g=\sqrt{\frac{1-t}{t}}$, the robustness of the state can be enhanced. We also consider the effect of limited coupling efficiency $\eta_c$ out of the PDC sources (i.e. out of the non-linear crystals).}\label{fig.sourceamp}
\end{figure}

\section{Amplification at the source}
\label{sec.ampsource}

In this section, we will present a scheme for realising the optimal states presented above. The main characteristic of our scheme is that it uses a heralded noiseless amplifier. In particular, we consider a standard parametric down-conversion (PDC) source and place the amplifier in one of the output modes of the PDC (see \figref{fig.sourceamp}). After the amplification, the PDC output modes are recombined on a 50/50 beam-splitter. The PDC source produces a two-mode squeezed state of the form 
\ba
 \ket{\psi}=\sqrt{1-\lambda^2}\sum_{n}\lambda^n \ket{n}\ket{n}
\ea
where $\lambda$ is the squeezing parameter. In the case without amplification, we obtain an output state (after the beam-splitter) of the form $\ket{00} + \lambda(\ket{20} + \ket{02}) + O(\lambda^2)$. Although this state has the desired form, the single parameter $\lambda$ does not allow us to recover exactly the optimal states of the previous section, i.e.~eqs.~\eqnref{eq.optstate_symt}, \eqnref{eq.optstate_symd}. Therefore, we introduce the amplifier on one arm of the PDC source.

The amplifier is shown on the right-hand side of \figref{fig.sourceamp} and is based on Ref.~\cite{ralph2009}. It makes use of an ancilla, a heralded single photon, created by PDC followed by detector $D_0$. The single photon is split on a beam splitter with variable transmission $t$ and reflectivity $r=1-t$. Typically we consider the regime $t<<1$. Next, the transmitted signal interferes with the input beam on a 50/50 beam splitter, the outputs of which are finally measured by two additional single-photon detectors ($D_1$ and $D_2$). Successful amplification requires exactly one of these two detectors to click. Thus a complete amplification is heralded by two detector clicks (either $\{D_0,D_1\}$ or $\{D_0,D_2\}$), taking into account the successful preparation of the single photon ancilla. Importantly, it is only upon successful amplification that the state is transmitted to Alice and Bob. If amplification is not successful, the state preparation must be repeated. This introduces no loophole, since the choice of local measurement settings for Alice and Bob can be done at a later time, and only upon successful operation of the source.

\begin{figure}
\includegraphics[width=.48\textwidth]{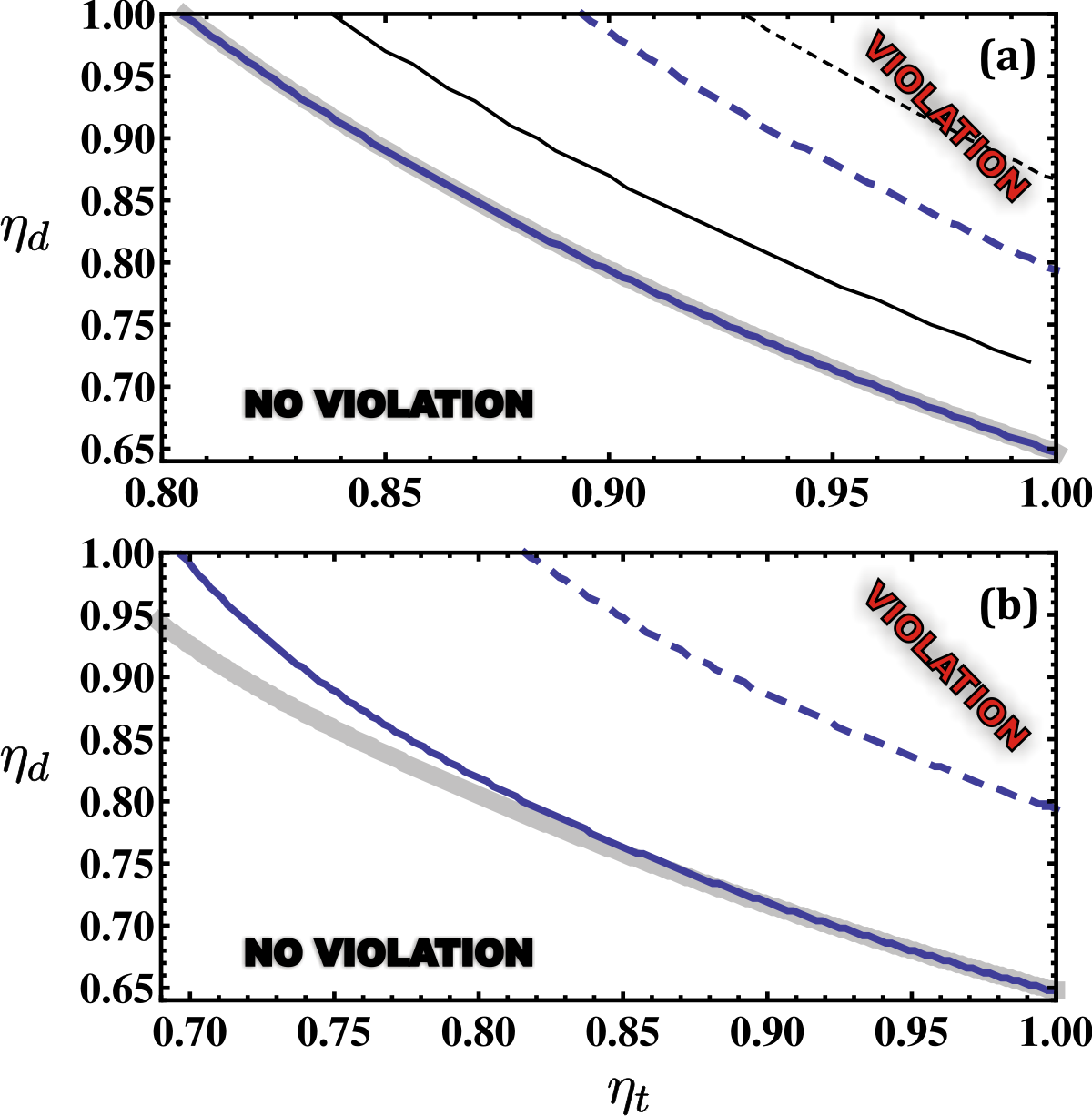}
\caption{(Colour online) Amplification at the source. The graphs show the region of Bell violations as a function of the transmission $\eta_t$ and the detection efficiency $\eta_d$, for symmetric \textbf{(a)} and asymmetric \textbf{(b)} setups. We consider two coupling effiencies: $\eta_c=1$ (solid curves) and $\eta_c=0.9$ (dashed curves). The CHSH inequality is violated for all combinations of transmission and detection efficiencies lying above the curves. As a reference we plot the regions obtained with optimal states in the 2-photon subspace (fat gray curves) as well as those of the scheme of Ref.~\cite{cavalcanti2011} (thin black curves) over which we get a significant improvement.}\label{fig.regions_sourceamp} 
\end{figure}

To understand how the amplifier modifies our entangled state, it is enough to recall how it acts on a weak coherent state of the form $\ket{0}+\alpha \ket{1}$. The output state will be of the form $\sqrt{t}\ket{0}+\alpha \sqrt{r} \ket{1}$, hence the effect is a noiseless amplification with gain $g=\sqrt{\frac{r}{t}}$. Indeed $g>1$ when $r>t$. Note that the success probability of the amplification is $~t$; hence the larger the gain, the smaller the probability of success. Now, in our case, a successful application of the amplifier to our initial two-mode squeezed state gives an output state (after the beam-splitter)
\ba
 \ket{\psi} = \sqrt{t} \ket{00} + \frac{\lambda \sqrt{r}}{\sqrt{2}} ( \ket{20} + \ket{02})
\ea 
which has the desired form. Therefore, by choosing appropriate values of squeezing $\lambda$ and gain $g$, we can produce the optimal states of the previous section. By including an amplifier in the source, we are thus able to produce states tolerating lower transmission and detection efficiencies as compared to Ref.~\cite{cavalcanti2011} and which are optimal in the 2-photon subspace. It is important to note that the complexity of our source is comparable to that of Ref.~\cite{cavalcanti2011}, in the sense that both schemes make use of two PDC sources. Moreover, the amplifier has been experimentally demonstrated \cite{xiang2010,ferreyrol2010}. Note however that for the asymmetric case, our scheme does not allow one to create the optimal state for transmission \eqref{eq.optstate_asymt} which requires imbalanced $\ket{20}$ and $\ket{02}$ terms.

Fig.~\ref{fig.regions_sourceamp} presents the results, by showing the threshold values for transmission and detection effciency leading to a violation of the CHSH inequality. From the perspective of experimental implementations, it is essential to account for losses due to imperfect coupling out of the PDC into the transmission channels. Couplings of the order of $80-90\%$ have been experimentally achieved \cite{pittman2005}. We have taken into account such coupling losses for both PDC sources. \figref{fig.regions_sourceamp} presents results for the cases $\eta_c=1$ and $\eta_c=0.9$. Moreover, one should also take into account the effect of higher-order contributions from the squeezing. We have checked that considering terms containing up to 6 photons, our results remain unchanged. Finally we note that for small values of $\lambda$ and $t$, the detection losses in the amplifier only affect the rate of successful amplification. As long as this rate is significantly larger than the technical noise (e.g.~dark counts), our Bell test will not be affected.

\section{Amplification by the parties}
\label{sec.ampparties}

In this section we change gears and consider a scenario where Alice and Bob amplify their systems locally, before performing their local measurements for the Bell test (see \figref{fig.setup}b). Importantly, it is only when the amplification succeeds that Alice and Bob proceed with the Bell test and choose a measurement basis. At the final stage of the Bell test, Alice and Bob will compare their statistics and keep only those events in which amplification succeeded for both parties. The amplification can thus be seen as part of the state-preparation process. We shall see that in this way, the detrimental effect of losses can be compensated, resulting in a higher and more robust Bell inequality violation. Note that a recent proposal for the implementation of device-independent quantum key distribution~\cite{gisin2010} is based on a similar use of amplifiers. Local amplification by the parties was also applied to a single-photon Bell test in Ref.~\cite{brask2012}.

Here we consider a symmetric configuration, and use a type of amplifiers described in \cite{marek2010}. The amplifier can be conveniently described by an operator of the form
\be
\hat{G}(g)=(g-2)\ah^\dag \ah + \ah\ah^\dag = (g-1)\hat{n}+1 ,
\ee
where $\ah$, $\ah^\dag$ are the annihilation and creation operators respectively, $\hat{n}=\ah^\dag\ah$ and $g$ is again the gain of the amplifier. This amplifier acts on Fock states as
\be
\hat{G}(g)\ket{n} = [(g-1)\hat{n}+1]\ket{n}.
\ee
The particular case $g=2$ was experimentally demonstrated in Ref.~\cite{zavatta2010} and corresponds to sequential addition and subtraction of a photon.

\begin{figure}
\includegraphics[width=.48\textwidth]{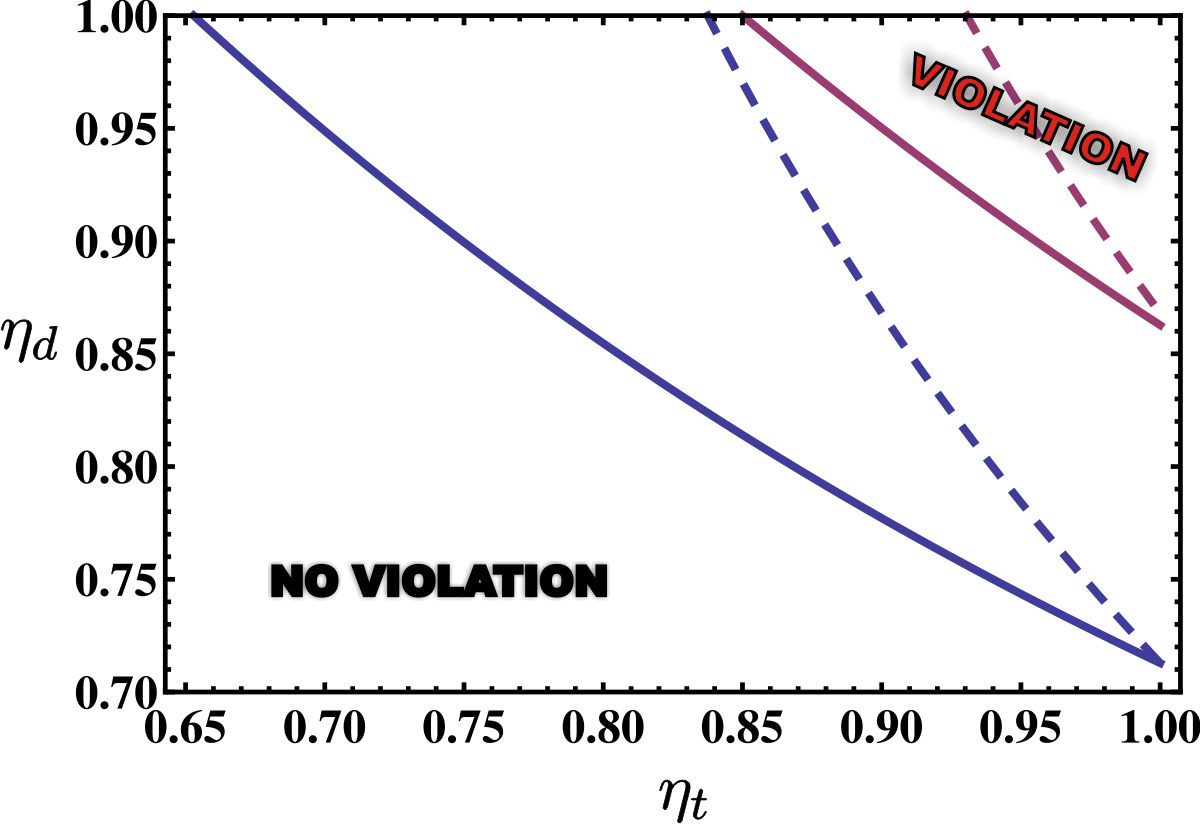}
\caption{Amplification by the parties. Regions of Bell violation as a function of the transmission $\eta_t$ and the detection efficiency $\eta_d$ for the state $\ket{\psi_2}$. The solid curves represent the case in which the local amplifiers have gain $g=2$. We obtain a significant improvement over the case without amplification \cite{cavalcanti2011} (dashed curves). For each case, we consider two coupling efficiencies, $\eta_c=1$ (lower curve) and $\eta_c=0.9$ (upper curve). Note that in the case with amplifiers, we consider all PDC sources (including those used for amplification) to have the same coupling efficiency.}\label{regions_partamp}
\end{figure}

We examine the effect of amplification on an entangled state of the form $\ket{\psi_2} = \frac{1}{\sqrt{2}}(\ket{02}+\ket{20})$, as in Ref.~\cite{cavalcanti2011}. The state arriving at Alice's and Bob's locations after lossy transmission ($\eta_t$) is then given by
\be
\rho = \eta_t^2\ket{\psi_2}\bra{\psi_2} + \eta_t(1-\eta_t)\rho_1+(1-\eta_t)^2\rho_0 ,
\ee
with
\be
\rho_1 = \ketbra{01}{01}+\ketbra{10}{10} , \hspace{0.5cm} \rho_0 = \ketbra{00}{00} .
\ee
If no amplification is performed, the minimum transmissivity needed for a CHSH violation is $\ceta_t=0.84$ (see \figref{fig.regions_sourceamp}) \cite{cavalcanti2011}. However, if Alice and Bob now apply $\hat{G}(g)$ before proceeding with their measurements, the state is transformed to
\begin{align}
\rho_g = & (\hat{G}\otimes\hat{G})\rho(\hat{G^\dag}\otimes\hat{G^\dag}) \nonumber \\
= & \frac{1}{N} [ (2g-1)^2 \eta_t^2\ket{\psi_2}\bra{\psi_2} + g^2 \eta_t(1-\eta_t)\rho_1 \\
& + (1-\eta_t)^2\rho_0 ]  \nonumber ,
\end{align}
where $N=1 - 4 g \eta_t^2 + 2 g^2 \eta_t (1 + \eta_t)$ is a normalization factor. From the last expression it is clear that the $\ket{\psi_2}$-component, i.e.~the original state, is increased with respect to the other components. The critical transmission efficiencies in the two instances of the amplifier, $g=2$ and $g=3$ become $\ceta_t = 0.62$ and $\ceta_t = 0.5$ respectively. Thus, we see that the amplification significantly helps in decreasing the threshold. In \figref{regions_partamp}, we show the region of parameters $\eta_d$, $\eta_t$ for which a violation of CHSH can be found. We stress that, using filters, one can obtain a violation of CHSH for parameters comparable to those reported in recent experiments. For instance, our scheme achieves CHSH violations with values $\eta_d= \eta_t=0.8$ (note that coupling and transmission losses are equivalent for the symmetric setups we consider here). Ref.~\cite{smith2012} recently reported similar values, reaching a total efficiency (including coupling, transmission and detection) of $\sim 62\%$.

One may also wonder about combining local amplifiers with the scheme presented above, in which an amplifier is used in the source. We have however checked that no improvement is obtained in this case. For completeness, we have also considered the case where both Alice and Bob apply the filter several times. The critical efficiency $\eta_t$ is found to decrease exponentially with the number of successive filters (see Appendix).

\section{Conclusion}
\label{sec.conclusion}

We have presented continuous variable Bell tests based on hybrid measurements and heralded amplifiers, which provide significant improvements in tolerance to transmission losses and detection efficiency. We discussed two types of schemes, in which the amplification is performed either at the source or at Alice and Bob's labs, prior to the local measurements. 

We believe that both of our schemes are relevant from the practical point of view. First, in the case of an amplification at the source, the complexity of the setup is basically similar to that of the scheme of Ref. \cite{cavalcanti2011}. The advantage in terms of both transmission losses and detection efficiency is nevertheless significant. Second, in the case of local amplification, the threshold values can be further reduced, approaching values recently achieved experimentally. However, the price to pay here is an increased complexity of the setup due to the requirement of additional photons. Moreover, this scheme may find applications in the implementation of device-independent tasks, such as quantum key distribution.

\textbf{Acknowledgements.} We would like to acknowledge helpful discussions with T.~Fritz and A.~Ac\'in. J.~B.~B. was funded by the Carlsberg Foundation and ERC starting grant PERCENT, N.~B.~by the UK EPSRC and the EU project DIQIP, A.~L.~by the ERC starting grant PERCENT, and D.~C.~acknowledges the National Research Foundation, the Ministry of Education (Singapore), and the PVE-CAPES program (Brazil).

\section{Appendix}

\begin{figure}
\includegraphics[width=.48\textwidth]{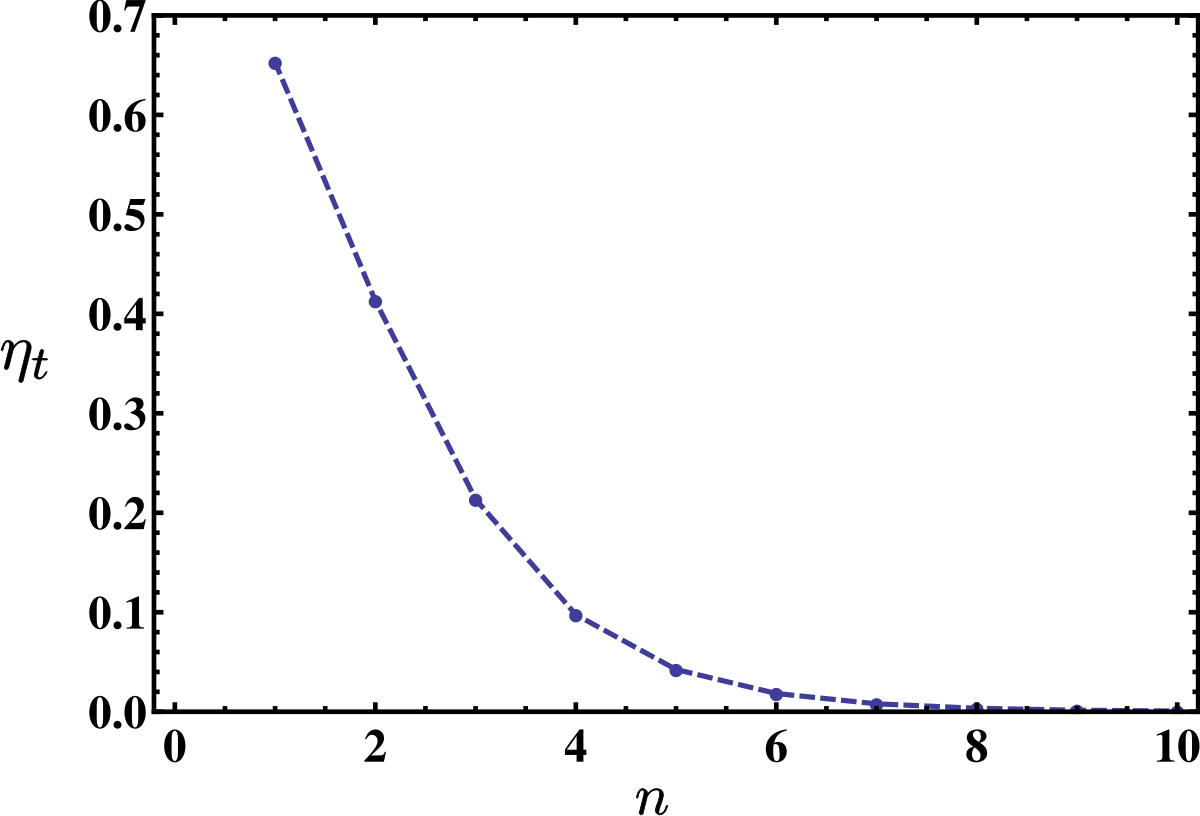}
\caption{Multiple amplification by the parties. Critical transmission as a function of the number of applications of the filter $\hat{G}(g)$, with gain $g=2$. The initial state is $\ket{\psi_2}$.}\label{fig.multiamp}
\end{figure}

We consider the case where both Alice and Bob apply the filter $\hat{G}(g)$ several times before performing their local measurements. We find that the critical $\eta_t$ decreases exponentially with the number of amplifiers. As a matter of comparison, in \figref{fig.multiamp} we plot the critical transmission required for a CHSH violation as a function of the number of times the filter $\hat{G}(2)$ is applied by Alice and Bob, in the case of the state $\ket{\psi_2}$ without vacuum component. Note that for three applications of the filter the critical transmission already drops to $\ceta_t=0.2$.

\bibliography{hybridbellamp}

\begin{thebibliography}{36}%
\makeatletter
\providecommand \@ifxundefined [1]{%
 \@ifx{#1\undefined}
}%
\providecommand \@ifnum [1]{%
 \ifnum #1\expandafter \@firstoftwo
 \else \expandafter \@secondoftwo
 \fi
}%
\providecommand \@ifx [1]{%
 \ifx #1\expandafter \@firstoftwo
 \else \expandafter \@secondoftwo
 \fi
}%
\providecommand \natexlab [1]{#1}%
\providecommand \enquote  [1]{``#1''}%
\providecommand \bibnamefont  [1]{#1}%
\providecommand \bibfnamefont [1]{#1}%
\providecommand \citenamefont [1]{#1}%
\providecommand \href@noop [0]{\@secondoftwo}%
\providecommand \href [0]{\begingroup \@sanitize@url \@href}%
\providecommand \@href[1]{\@@startlink{#1}\@@href}%
\providecommand \@@href[1]{\endgroup#1\@@endlink}%
\providecommand \@sanitize@url [0]{\catcode `\\12\catcode `\$12\catcode
  `\&12\catcode `\#12\catcode `\^12\catcode `\_12\catcode `\%12\relax}%
\providecommand \@@startlink[1]{}%
\providecommand \@@endlink[0]{}%
\providecommand \url  [0]{\begingroup\@sanitize@url \@url }%
\providecommand \@url [1]{\endgroup\@href {#1}{\urlprefix }}%
\providecommand \urlprefix  [0]{URL }%
\providecommand \Eprint [0]{\href }%
\providecommand \doibase [0]{http://dx.doi.org/}%
\providecommand \selectlanguage [0]{\@gobble}%
\providecommand \bibinfo  [0]{\@secondoftwo}%
\providecommand \bibfield  [0]{\@secondoftwo}%
\providecommand \translation [1]{[#1]}%
\providecommand \BibitemOpen [0]{}%
\providecommand \bibitemStop [0]{}%
\providecommand \bibitemNoStop [0]{.\EOS\space}%
\providecommand \EOS [0]{\spacefactor3000\relax}%
\providecommand \BibitemShut  [1]{\csname bibitem#1\endcsname}%
\let\auto@bib@innerbib\@empty
\bibitem [{\citenamefont {Bell}(1964)}]{bell1964}%
  \BibitemOpen
  \bibfield  {author} {\bibinfo {author} {\bibfnamefont {J.}~\bibnamefont
  {Bell}},\ }\href@noop {} {\bibfield  {journal} {\bibinfo  {journal}
  {Physics}\ }\textbf {\bibinfo {volume} {1}},\ \bibinfo {pages} {195}
  (\bibinfo {year} {1964})}\BibitemShut {NoStop}%
\bibitem [{\citenamefont {Buhrman}\ \emph {et~al.}(2010)\citenamefont
  {Buhrman}, \citenamefont {Cleve}, \citenamefont {Massar},\ and\ \citenamefont
  {de~Wolf}}]{buhrman2010}%
  \BibitemOpen
  \bibfield  {author} {\bibinfo {author} {\bibfnamefont {H.}~\bibnamefont
  {Buhrman}}, \bibinfo {author} {\bibfnamefont {R.}~\bibnamefont {Cleve}},
  \bibinfo {author} {\bibfnamefont {S.}~\bibnamefont {Massar}}, \ and\ \bibinfo
  {author} {\bibfnamefont {R.}~\bibnamefont {de~Wolf}},\ }\href {\doibase
  10.1103/RevModPhys.82.665} {\bibfield  {journal} {\bibinfo  {journal} {Rev.
  Mod. Phys.}\ }\textbf {\bibinfo {volume} {82}},\ \bibinfo {pages} {665}
  (\bibinfo {year} {2010})}\BibitemShut {NoStop}%
\bibitem [{\citenamefont {Acin}\ \emph {et~al.}(2007)\citenamefont {Acin},
  \citenamefont {Brunner}, \citenamefont {Gisin}, \citenamefont {Massar},
  \citenamefont {Pironio},\ and\ \citenamefont {Scarani}}]{acin2007}%
  \BibitemOpen
  \bibfield  {author} {\bibinfo {author} {\bibfnamefont {A.}~\bibnamefont
  {Acin}}, \bibinfo {author} {\bibfnamefont {N.}~\bibnamefont {Brunner}},
  \bibinfo {author} {\bibfnamefont {N.}~\bibnamefont {Gisin}}, \bibinfo
  {author} {\bibfnamefont {S.}~\bibnamefont {Massar}}, \bibinfo {author}
  {\bibfnamefont {S.}~\bibnamefont {Pironio}}, \ and\ \bibinfo {author}
  {\bibfnamefont {V.}~\bibnamefont {Scarani}},\ }\href@noop {} {\bibfield
  {journal} {\bibinfo  {journal} {Phys. Rev. Lett.}\ }\textbf {\bibinfo
  {volume} {98}},\ \bibinfo {pages} {230501} (\bibinfo {year}
  {2007})}\BibitemShut {NoStop}%
\bibitem [{\citenamefont {H{\"a}nggi}\ and\ \citenamefont
  {Renner}(2010)}]{hanggi2010}%
  \BibitemOpen
  \bibfield  {author} {\bibinfo {author} {\bibfnamefont {E.}~\bibnamefont
  {H{\"a}nggi}}\ and\ \bibinfo {author} {\bibfnamefont {R.}~\bibnamefont
  {Renner}},\ }\href@noop {} {\bibfield  {journal} {\bibinfo  {journal}
  {e-print arXiv:1009.1833}\ } (\bibinfo {year} {2010})}\BibitemShut {NoStop}%
\bibitem [{\citenamefont {Masanes}\ \emph {et~al.}(2011)\citenamefont
  {Masanes}, \citenamefont {Pironio},\ and\ \citenamefont
  {Ac{\'\i}n}}]{masanes2011}%
  \BibitemOpen
  \bibfield  {author} {\bibinfo {author} {\bibfnamefont {L.}~\bibnamefont
  {Masanes}}, \bibinfo {author} {\bibfnamefont {S.}~\bibnamefont {Pironio}}, \
  and\ \bibinfo {author} {\bibfnamefont {A.}~\bibnamefont {Ac{\'\i}n}},\
  }\href@noop {} {\bibfield  {journal} {\bibinfo  {journal} {Nat. Commun.}\
  }\textbf {\bibinfo {volume} {2}},\ \bibinfo {pages} {238} (\bibinfo {year}
  {2011})}\BibitemShut {NoStop}%
\bibitem [{\citenamefont {Pironio}\ \emph {et~al.}(2010)\citenamefont {Pironio}
  \emph {et~al.}}]{pironio2010}%
  \BibitemOpen
  \bibfield  {author} {\bibinfo {author} {\bibfnamefont {S.}~\bibnamefont
  {Pironio}} \emph {et~al.},\ }\href@noop {} {\bibfield  {journal} {\bibinfo
  {journal} {Nature}\ }\textbf {\bibinfo {volume} {464}},\ \bibinfo {pages}
  {1021} (\bibinfo {year} {2010})}\BibitemShut {NoStop}%
\bibitem [{\citenamefont {Colbeck}\ and\ \citenamefont
  {Kent}(2011)}]{colbeck2011}%
  \BibitemOpen
  \bibfield  {author} {\bibinfo {author} {\bibfnamefont {R.}~\bibnamefont
  {Colbeck}}\ and\ \bibinfo {author} {\bibfnamefont {A.}~\bibnamefont {Kent}},\
  }\href@noop {} {\bibfield  {journal} {\bibinfo  {journal} {J. Phys. A}\
  }\textbf {\bibinfo {volume} {44}},\ \bibinfo {pages} {095305} (\bibinfo
  {year} {2011})}\BibitemShut {NoStop}%
\bibitem [{\citenamefont {Bardyn}\ \emph {et~al.}(2009)\citenamefont {Bardyn},
  \citenamefont {Liew}, \citenamefont {Massar}, \citenamefont {McKague},\ and\
  \citenamefont {Scarani}}]{bardyn2009}%
  \BibitemOpen
  \bibfield  {author} {\bibinfo {author} {\bibfnamefont {C.-E.}\ \bibnamefont
  {Bardyn}}, \bibinfo {author} {\bibfnamefont {T.~C.~H.}\ \bibnamefont {Liew}},
  \bibinfo {author} {\bibfnamefont {S.}~\bibnamefont {Massar}}, \bibinfo
  {author} {\bibfnamefont {M.}~\bibnamefont {McKague}}, \ and\ \bibinfo
  {author} {\bibfnamefont {V.}~\bibnamefont {Scarani}},\ }\href {\doibase
  10.1103/PhysRevA.80.062327} {\bibfield  {journal} {\bibinfo  {journal} {Phys.
  Rev. A}\ }\textbf {\bibinfo {volume} {80}},\ \bibinfo {pages} {062327}
  (\bibinfo {year} {2009})}\BibitemShut {NoStop}%
\bibitem [{\citenamefont {Bancal}\ \emph {et~al.}(2011)\citenamefont {Bancal},
  \citenamefont {Gisin}, \citenamefont {Liang},\ and\ \citenamefont
  {Pironio}}]{bancal2011}%
  \BibitemOpen
  \bibfield  {author} {\bibinfo {author} {\bibfnamefont {J.-D.}\ \bibnamefont
  {Bancal}}, \bibinfo {author} {\bibfnamefont {N.}~\bibnamefont {Gisin}},
  \bibinfo {author} {\bibfnamefont {Y.-C.}\ \bibnamefont {Liang}}, \ and\
  \bibinfo {author} {\bibfnamefont {S.}~\bibnamefont {Pironio}},\ }\href
  {\doibase 10.1103/PhysRevLett.106.250404} {\bibfield  {journal} {\bibinfo
  {journal} {Phys. Rev. Lett.}\ }\textbf {\bibinfo {volume} {106}},\ \bibinfo
  {pages} {250404} (\bibinfo {year} {2011})}\BibitemShut {NoStop}%
\bibitem [{\citenamefont {Rabelo}\ \emph {et~al.}(2011)\citenamefont {Rabelo},
  \citenamefont {Ho}, \citenamefont {Cavalcanti}, \citenamefont {Brunner},\
  and\ \citenamefont {Scarani}}]{rabelo2011}%
  \BibitemOpen
  \bibfield  {author} {\bibinfo {author} {\bibfnamefont {R.}~\bibnamefont
  {Rabelo}}, \bibinfo {author} {\bibfnamefont {M.}~\bibnamefont {Ho}}, \bibinfo
  {author} {\bibfnamefont {D.}~\bibnamefont {Cavalcanti}}, \bibinfo {author}
  {\bibfnamefont {N.}~\bibnamefont {Brunner}}, \ and\ \bibinfo {author}
  {\bibfnamefont {V.}~\bibnamefont {Scarani}},\ }\href {\doibase
  10.1103/PhysRevLett.107.050502} {\bibfield  {journal} {\bibinfo  {journal}
  {Phys. Rev. Lett.}\ }\textbf {\bibinfo {volume} {107}},\ \bibinfo {pages}
  {050502} (\bibinfo {year} {2011})}\BibitemShut {NoStop}%
\bibitem [{\citenamefont {Aspect}(1999)}]{aspect1999}%
  \BibitemOpen
  \bibfield  {author} {\bibinfo {author} {\bibfnamefont {A.}~\bibnamefont
  {Aspect}},\ }\href@noop {} {\bibfield  {journal} {\bibinfo  {journal}
  {Nature}\ }\textbf {\bibinfo {volume} {398}},\ \bibinfo {pages} {189}
  (\bibinfo {year} {1999})}\BibitemShut {NoStop}%
\bibitem [{\citenamefont {Pearle}(1970)}]{pearle1970}%
  \BibitemOpen
  \bibfield  {author} {\bibinfo {author} {\bibfnamefont {P.~M.}\ \bibnamefont
  {Pearle}},\ }\href {\doibase 10.1103/PhysRevD.2.1418} {\bibfield  {journal}
  {\bibinfo  {journal} {Phys. Rev. D}\ }\textbf {\bibinfo {volume} {2}},\
  \bibinfo {pages} {1418} (\bibinfo {year} {1970})}\BibitemShut {NoStop}%
\bibitem [{\citenamefont {Rowe}\ \emph {et~al.}(2001)\citenamefont {Rowe},
  \citenamefont {Kielpinski}, \citenamefont {Meyer}, \citenamefont {Sackett},
  \citenamefont {Itano}, \citenamefont {Monroe},\ and\ \citenamefont
  {Wineland}}]{rowe2001}%
  \BibitemOpen
  \bibfield  {author} {\bibinfo {author} {\bibfnamefont {M.~A.}\ \bibnamefont
  {Rowe}}, \bibinfo {author} {\bibfnamefont {D.}~\bibnamefont {Kielpinski}},
  \bibinfo {author} {\bibfnamefont {V.}~\bibnamefont {Meyer}}, \bibinfo
  {author} {\bibfnamefont {C.~A.}\ \bibnamefont {Sackett}}, \bibinfo {author}
  {\bibfnamefont {W.~M.}\ \bibnamefont {Itano}}, \bibinfo {author}
  {\bibfnamefont {C.}~\bibnamefont {Monroe}}, \ and\ \bibinfo {author}
  {\bibfnamefont {D.~J.}\ \bibnamefont {Wineland}},\ }\href {\doibase
  10.1038/35057215} {\bibfield  {journal} {\bibinfo  {journal} {Nature}\
  }\textbf {\bibinfo {volume} {409}},\ \bibinfo {pages} {791} (\bibinfo {year}
  {2001})}\BibitemShut {NoStop}%
\bibitem [{\citenamefont {Gisin}\ \emph {et~al.}(2010)\citenamefont {Gisin},
  \citenamefont {Pironio},\ and\ \citenamefont {Sangouard}}]{gisin2010}%
  \BibitemOpen
  \bibfield  {author} {\bibinfo {author} {\bibfnamefont {N.}~\bibnamefont
  {Gisin}}, \bibinfo {author} {\bibfnamefont {S.}~\bibnamefont {Pironio}}, \
  and\ \bibinfo {author} {\bibfnamefont {N.}~\bibnamefont {Sangouard}},\
  }\href@noop {} {\bibfield  {journal} {\bibinfo  {journal} {Phys. Rev. Lett.}\
  }\textbf {\bibinfo {volume} {105}},\ \bibinfo {pages} {70501} (\bibinfo
  {year} {2010})}\BibitemShut {NoStop}%
\bibitem [{\citenamefont {Gilchrist}\ \emph {et~al.}(1998)\citenamefont
  {Gilchrist}, \citenamefont {Deuar},\ and\ \citenamefont
  {Reid}}]{gilchrist1998}%
  \BibitemOpen
  \bibfield  {author} {\bibinfo {author} {\bibfnamefont {A.}~\bibnamefont
  {Gilchrist}}, \bibinfo {author} {\bibfnamefont {P.}~\bibnamefont {Deuar}}, \
  and\ \bibinfo {author} {\bibfnamefont {M.~D.}\ \bibnamefont {Reid}},\ }\href
  {\doibase 10.1103/PhysRevLett.80.3169} {\bibfield  {journal} {\bibinfo
  {journal} {Phys. Rev. Lett.}\ }\textbf {\bibinfo {volume} {80}},\ \bibinfo
  {pages} {3169} (\bibinfo {year} {1998})}\BibitemShut {NoStop}%
\bibitem [{\citenamefont {Banaszek}\ and\ \citenamefont
  {W\'odkiewicz}(1998)}]{banaszek1998}%
  \BibitemOpen
  \bibfield  {author} {\bibinfo {author} {\bibfnamefont {K.}~\bibnamefont
  {Banaszek}}\ and\ \bibinfo {author} {\bibfnamefont {K.}~\bibnamefont
  {W\'odkiewicz}},\ }\href {\doibase 10.1103/PhysRevA.58.4345} {\bibfield
  {journal} {\bibinfo  {journal} {Phys. Rev. A}\ }\textbf {\bibinfo {volume}
  {58}},\ \bibinfo {pages} {4345} (\bibinfo {year} {1998})}\BibitemShut
  {NoStop}%
\bibitem [{\citenamefont {Garc\'\i{}a-Patr\'on}\ \emph
  {et~al.}(2004)\citenamefont {Garc\'\i{}a-Patr\'on}, \citenamefont
  {Fiur\'a\ifmmode~\check{s}\else \v{s}\fi{}ek}, \citenamefont {Cerf},
  \citenamefont {Wenger}, \citenamefont {Tualle-Brouri},\ and\ \citenamefont
  {Grangier}}]{garcia-patron2004}%
  \BibitemOpen
  \bibfield  {author} {\bibinfo {author} {\bibfnamefont {R.}~\bibnamefont
  {Garc\'\i{}a-Patr\'on}}, \bibinfo {author} {\bibfnamefont {J.}~\bibnamefont
  {Fiur\'a\ifmmode~\check{s}\else \v{s}\fi{}ek}}, \bibinfo {author}
  {\bibfnamefont {N.~J.}\ \bibnamefont {Cerf}}, \bibinfo {author}
  {\bibfnamefont {J.}~\bibnamefont {Wenger}}, \bibinfo {author} {\bibfnamefont
  {R.}~\bibnamefont {Tualle-Brouri}}, \ and\ \bibinfo {author} {\bibfnamefont
  {P.}~\bibnamefont {Grangier}},\ }\href@noop {} {\bibfield  {journal}
  {\bibinfo  {journal} {Phys. Rev. Lett.}\ }\textbf {\bibinfo {volume} {93}},\
  \bibinfo {pages} {130409} (\bibinfo {year} {2004})}\BibitemShut {NoStop}%
\bibitem [{\citenamefont {Nha}\ and\ \citenamefont
  {Carmichael}(2004)}]{nha2004}%
  \BibitemOpen
  \bibfield  {author} {\bibinfo {author} {\bibfnamefont {H.}~\bibnamefont
  {Nha}}\ and\ \bibinfo {author} {\bibfnamefont {H.~J.}\ \bibnamefont
  {Carmichael}},\ }\href@noop {} {\bibfield  {journal} {\bibinfo  {journal}
  {Phys. Rev. Lett.}\ }\textbf {\bibinfo {volume} {93}},\ \bibinfo {pages}
  {020401} (\bibinfo {year} {2004})}\BibitemShut {NoStop}%
\bibitem [{\citenamefont {Cavalcanti}\ \emph {et~al.}(2011)\citenamefont
  {Cavalcanti}, \citenamefont {Brunner}, \citenamefont {Skrzypczyk},
  \citenamefont {Salles},\ and\ \citenamefont {Scarani}}]{cavalcanti2011}%
  \BibitemOpen
  \bibfield  {author} {\bibinfo {author} {\bibfnamefont {D.}~\bibnamefont
  {Cavalcanti}}, \bibinfo {author} {\bibfnamefont {N.}~\bibnamefont {Brunner}},
  \bibinfo {author} {\bibfnamefont {P.}~\bibnamefont {Skrzypczyk}}, \bibinfo
  {author} {\bibfnamefont {A.}~\bibnamefont {Salles}}, \ and\ \bibinfo {author}
  {\bibfnamefont {V.}~\bibnamefont {Scarani}},\ }\href@noop {} {\bibfield
  {journal} {\bibinfo  {journal} {Phys. Rev. A}\ }\textbf {\bibinfo {volume}
  {84}},\ \bibinfo {pages} {022105} (\bibinfo {year} {2011})}\BibitemShut
  {NoStop}%
\bibitem [{\citenamefont {V\'ertesi}\ \emph {et~al.}(2010)\citenamefont
  {V\'ertesi}, \citenamefont {Pironio},\ and\ \citenamefont
  {Brunner}}]{vertesi2010}%
  \BibitemOpen
  \bibfield  {author} {\bibinfo {author} {\bibfnamefont {T.}~\bibnamefont
  {V\'ertesi}}, \bibinfo {author} {\bibfnamefont {S.}~\bibnamefont {Pironio}},
  \ and\ \bibinfo {author} {\bibfnamefont {N.}~\bibnamefont {Brunner}},\ }\href
  {\doibase 10.1103/PhysRevLett.104.060401} {\bibfield  {journal} {\bibinfo
  {journal} {Phys. Rev. Lett.}\ }\textbf {\bibinfo {volume} {104}},\ \bibinfo
  {pages} {060401} (\bibinfo {year} {2010})}\BibitemShut {NoStop}%
\bibitem [{\citenamefont {Quintino}\ \emph {et~al.}(2011)\citenamefont
  {Quintino}, \citenamefont {Araújo}, \citenamefont {Cavalcanti}, \citenamefont
  {Santos},\ and\ \citenamefont {Cunha}}]{quintino2011}%
  \BibitemOpen
  \bibfield  {author} {\bibinfo {author} {\bibfnamefont {M.~T.}\ \bibnamefont
  {Quintino}}, \bibinfo {author} {\bibfnamefont {M.}~\bibnamefont {Araújo}},
  \bibinfo {author} {\bibfnamefont {D.}~\bibnamefont {Cavalcanti}}, \bibinfo
  {author} {\bibfnamefont {M.~F.}\ \bibnamefont {Santos}}, \ and\ \bibinfo
  {author} {\bibfnamefont {M.~T.}\ \bibnamefont {Cunha}},\ }\href@noop {}
  {\bibfield  {journal} {\bibinfo  {journal} {e-print arXiv:1106.2486v2}\ }
  (\bibinfo {year} {2011})}\BibitemShut {NoStop}%
\bibitem [{\citenamefont {Ara{\'u}jo}\ \emph {et~al.}(2011)\citenamefont
  {Ara{\'u}jo} \emph {et~al.}}]{araujo2011}%
  \BibitemOpen
  \bibfield  {author} {\bibinfo {author} {\bibfnamefont {M.}~\bibnamefont
  {Ara{\'u}jo}} \emph {et~al.},\ }\href@noop {} {\bibfield  {journal} {\bibinfo
   {journal} {e-print arXiv:1112.1719}\ } (\bibinfo {year} {2011})}\BibitemShut
  {NoStop}%
\bibitem [{\citenamefont {Sangouard}\ \emph {et~al.}(2011)\citenamefont
  {Sangouard}, \citenamefont {Bancal}, \citenamefont {Gisin}, \citenamefont
  {Rosenfeld}, \citenamefont {Sekatski}, \citenamefont {Weber},\ and\
  \citenamefont {Weinfurter}}]{sangouard2011}%
  \BibitemOpen
  \bibfield  {author} {\bibinfo {author} {\bibfnamefont {N.}~\bibnamefont
  {Sangouard}}, \bibinfo {author} {\bibfnamefont {J.-D.}\ \bibnamefont
  {Bancal}}, \bibinfo {author} {\bibfnamefont {N.}~\bibnamefont {Gisin}},
  \bibinfo {author} {\bibfnamefont {W.}~\bibnamefont {Rosenfeld}}, \bibinfo
  {author} {\bibfnamefont {P.}~\bibnamefont {Sekatski}}, \bibinfo {author}
  {\bibfnamefont {M.}~\bibnamefont {Weber}}, \ and\ \bibinfo {author}
  {\bibfnamefont {H.}~\bibnamefont {Weinfurter}},\ }\href {\doibase
  10.1103/PhysRevA.84.052122} {\bibfield  {journal} {\bibinfo  {journal} {Phys.
  Rev. A}\ }\textbf {\bibinfo {volume} {84}},\ \bibinfo {pages} {052122}
  (\bibinfo {year} {2011})}\BibitemShut {NoStop}%
\bibitem [{\citenamefont {Chaves}\ and\ \citenamefont
  {Brask}(2011)}]{chaves2011}%
  \BibitemOpen
  \bibfield  {author} {\bibinfo {author} {\bibfnamefont {R.}~\bibnamefont
  {Chaves}}\ and\ \bibinfo {author} {\bibfnamefont {J.~B.}\ \bibnamefont
  {Brask}},\ }\href@noop {} {\bibfield  {journal} {\bibinfo  {journal} {Phys.
  Rev. A}\ }\textbf {\bibinfo {volume} {84}},\ \bibinfo {pages} {062110}
  (\bibinfo {year} {2011})}\BibitemShut {NoStop}%
\bibitem [{\citenamefont {Brunner}\ \emph {et~al.}(2007)\citenamefont
  {Brunner}, \citenamefont {Gisin}, \citenamefont {Scarani},\ and\
  \citenamefont {Simon}}]{brunner2007}%
  \BibitemOpen
  \bibfield  {author} {\bibinfo {author} {\bibfnamefont {N.}~\bibnamefont
  {Brunner}}, \bibinfo {author} {\bibfnamefont {N.}~\bibnamefont {Gisin}},
  \bibinfo {author} {\bibfnamefont {V.}~\bibnamefont {Scarani}}, \ and\
  \bibinfo {author} {\bibfnamefont {C.}~\bibnamefont {Simon}},\ }\href
  {\doibase 10.1103/PhysRevLett.98.220403} {\bibfield  {journal} {\bibinfo
  {journal} {Phys. Rev. Lett.}\ }\textbf {\bibinfo {volume} {98}},\ \bibinfo
  {pages} {220403} (\bibinfo {year} {2007})}\BibitemShut {NoStop}%
\bibitem [{\citenamefont {Cabello}\ and\ \citenamefont
  {Larsson}(2007)}]{cabello2007}%
  \BibitemOpen
  \bibfield  {author} {\bibinfo {author} {\bibfnamefont {A.}~\bibnamefont
  {Cabello}}\ and\ \bibinfo {author} {\bibfnamefont {J.-A.}\ \bibnamefont
  {Larsson}},\ }\href {\doibase 10.1103/PhysRevLett.98.220402} {\bibfield
  {journal} {\bibinfo  {journal} {Phys. Rev. Lett.}\ }\textbf {\bibinfo
  {volume} {98}},\ \bibinfo {pages} {220402} (\bibinfo {year}
  {2007})}\BibitemShut {NoStop}%
\bibitem [{\citenamefont {Ralph}\ and\ \citenamefont {Lund}(2009)}]{ralph2009}%
  \BibitemOpen
  \bibfield  {author} {\bibinfo {author} {\bibfnamefont {T.~C.}\ \bibnamefont
  {Ralph}}\ and\ \bibinfo {author} {\bibfnamefont {A.~P.}\ \bibnamefont
  {Lund}}\ }(\bibinfo  {publisher} {AIP, New York},\ \bibinfo {year} {2009})\
  pp.\ \bibinfo {pages} {155 (e--print arXiv:0809.0326)}\BibitemShut {NoStop}%
\bibitem [{\citenamefont {Marek}\ and\ \citenamefont
  {Filip}(2010)}]{marek2010}%
  \BibitemOpen
  \bibfield  {author} {\bibinfo {author} {\bibfnamefont {P.}~\bibnamefont
  {Marek}}\ and\ \bibinfo {author} {\bibfnamefont {R.}~\bibnamefont {Filip}},\
  }\href {\doibase 10.1103/PhysRevA.81.022302} {\bibfield  {journal} {\bibinfo
  {journal} {Phys. Rev. A}\ }\textbf {\bibinfo {volume} {81}},\ \bibinfo
  {pages} {022302} (\bibinfo {year} {2010})}\BibitemShut {NoStop}%
\bibitem [{\citenamefont {Xiang}\ \emph {et~al.}(2010)\citenamefont {Xiang},
  \citenamefont {Ralph}, \citenamefont {Lund}, \citenamefont {Walk},\ and\
  \citenamefont {Pryde}}]{xiang2010}%
  \BibitemOpen
  \bibfield  {author} {\bibinfo {author} {\bibfnamefont {G.}~\bibnamefont
  {Xiang}}, \bibinfo {author} {\bibfnamefont {T.}~\bibnamefont {Ralph}},
  \bibinfo {author} {\bibfnamefont {A.}~\bibnamefont {Lund}}, \bibinfo {author}
  {\bibfnamefont {N.}~\bibnamefont {Walk}}, \ and\ \bibinfo {author}
  {\bibfnamefont {G.}~\bibnamefont {Pryde}},\ }\href@noop {} {\bibfield
  {journal} {\bibinfo  {journal} {Nature Photonics}\ }\textbf {\bibinfo
  {volume} {4}},\ \bibinfo {pages} {316} (\bibinfo {year} {2010})}\BibitemShut
  {NoStop}%
\bibitem [{\citenamefont {Ferreyrol}\ \emph {et~al.}(2010)\citenamefont
  {Ferreyrol}, \citenamefont {Barbieri}, \citenamefont {Blandino},
  \citenamefont {Fossier}, \citenamefont {Tualle-Brouri},\ and\ \citenamefont
  {Grangier}}]{ferreyrol2010}%
  \BibitemOpen
  \bibfield  {author} {\bibinfo {author} {\bibfnamefont {F.}~\bibnamefont
  {Ferreyrol}}, \bibinfo {author} {\bibfnamefont {M.}~\bibnamefont {Barbieri}},
  \bibinfo {author} {\bibfnamefont {R.}~\bibnamefont {Blandino}}, \bibinfo
  {author} {\bibfnamefont {S.}~\bibnamefont {Fossier}}, \bibinfo {author}
  {\bibfnamefont {R.}~\bibnamefont {Tualle-Brouri}}, \ and\ \bibinfo {author}
  {\bibfnamefont {P.}~\bibnamefont {Grangier}},\ }\href {\doibase
  10.1103/PhysRevLett.104.123603} {\bibfield  {journal} {\bibinfo  {journal}
  {Phys. Rev. Lett.}\ }\textbf {\bibinfo {volume} {104}},\ \bibinfo {pages}
  {123603} (\bibinfo {year} {2010})}\BibitemShut {NoStop}%
\bibitem [{\citenamefont {Zavatta}\ \emph {et~al.}(2010)\citenamefont
  {Zavatta}, \citenamefont {Fiur{\'a}{\v{s}}ek},\ and\ \citenamefont
  {Bellini}}]{zavatta2010}%
  \BibitemOpen
  \bibfield  {author} {\bibinfo {author} {\bibfnamefont {A.}~\bibnamefont
  {Zavatta}}, \bibinfo {author} {\bibfnamefont {J.}~\bibnamefont
  {Fiur{\'a}{\v{s}}ek}}, \ and\ \bibinfo {author} {\bibfnamefont
  {M.}~\bibnamefont {Bellini}},\ }\href@noop {} {\bibfield  {journal} {\bibinfo
   {journal} {Nat. Photon.}\ }\textbf {\bibinfo {volume} {5}},\ \bibinfo
  {pages} {52} (\bibinfo {year} {2010})}\BibitemShut {NoStop}%
\bibitem [{\citenamefont {Usuga}\ \emph {et~al.}(2010)\citenamefont {Usuga}
  \emph {et~al.}}]{usuga2010}%
  \BibitemOpen
  \bibfield  {author} {\bibinfo {author} {\bibfnamefont {M.~A.}\ \bibnamefont
  {Usuga}} \emph {et~al.},\ }\href {http://dx.doi.org/10.1038/nphys1743}
  {\bibfield  {journal} {\bibinfo  {journal} {Nature Physics}\ }\textbf
  {\bibinfo {volume} {6}},\ \bibinfo {pages} {767} (\bibinfo {year}
  {2010})}\BibitemShut {NoStop}%
\bibitem [{\citenamefont {Clauser}\ \emph {et~al.}(1969)\citenamefont
  {Clauser}, \citenamefont {Horne}, \citenamefont {Shimony},\ and\
  \citenamefont {Holt}}]{chsh}%
  \BibitemOpen
  \bibfield  {author} {\bibinfo {author} {\bibfnamefont {J.~F.}\ \bibnamefont
  {Clauser}}, \bibinfo {author} {\bibfnamefont {M.~A.}\ \bibnamefont {Horne}},
  \bibinfo {author} {\bibfnamefont {A.}~\bibnamefont {Shimony}}, \ and\
  \bibinfo {author} {\bibfnamefont {R.~A.}\ \bibnamefont {Holt}},\ }\href
  {\doibase 10.1103/PhysRevLett.23.880} {\bibfield  {journal} {\bibinfo
  {journal} {Phys. Rev. Lett.}\ }\textbf {\bibinfo {volume} {23}},\ \bibinfo
  {pages} {880} (\bibinfo {year} {1969})}\BibitemShut {NoStop}%
\bibitem [{\citenamefont {Pittman}\ \emph {et~al.}(2005)\citenamefont
  {Pittman}, \citenamefont {Jacobs},\ and\ \citenamefont
  {Franson}}]{pittman2005}%
  \BibitemOpen
  \bibfield  {author} {\bibinfo {author} {\bibfnamefont {T.}~\bibnamefont
  {Pittman}}, \bibinfo {author} {\bibfnamefont {B.}~\bibnamefont {Jacobs}}, \
  and\ \bibinfo {author} {\bibfnamefont {J.}~\bibnamefont {Franson}},\
  }\href@noop {} {\bibfield  {journal} {\bibinfo  {journal} {Opt. Comm.}\
  }\textbf {\bibinfo {volume} {246}},\ \bibinfo {pages} {545} (\bibinfo {year}
  {2005})}\BibitemShut {NoStop}%
\bibitem [{\citenamefont {Brask}\ and\ \citenamefont
  {Chaves}(2012)}]{brask2012}%
  \BibitemOpen
  \bibfield  {author} {\bibinfo {author} {\bibfnamefont {J.~B.}\ \bibnamefont
  {Brask}}\ and\ \bibinfo {author} {\bibfnamefont {R.}~\bibnamefont {Chaves}},\
  }\href {http://arxiv.org/abs/1202.3049} {\bibfield  {journal} {\bibinfo
  {journal} {e-print arXiv:1202.3049}\ } (\bibinfo {year} {2012})}\BibitemShut
  {NoStop}%
\bibitem [{\citenamefont {Smith}\ \emph {et~al.}(2012)\citenamefont {Smith}
  \emph {et~al.}}]{smith2012}%
  \BibitemOpen
  \bibfield  {author} {\bibinfo {author} {\bibfnamefont {D.~H.}\ \bibnamefont
  {Smith}} \emph {et~al.},\ }\href {http://dx.doi.org/10.1038/ncomms1628}
  {\bibfield  {journal} {\bibinfo  {journal} {Nat. Commun.}\ }\textbf {\bibinfo
  {volume} {3}},\ \bibinfo {pages} {625} (\bibinfo {year} {2012})}\BibitemShut
  {NoStop}%
\end{thebibliography}%

\end{document}